# TeroSeek: An AI-Powered Knowledge Base and Retrieval Generation Platform for Terpenoid Research

Xu Kang, Siqi Jiang, Kangwei Xu, Jiahao Li, Ruibo Wu*

*School of Pharmaceutical Sciences, Sun Yat-sen University, Guangzhou 510006, P.R. China*

* **E-mail**: wurb3@mail.sysu.edu.cn

# Abstract

Terpenoids represent a pivotal class of natural products that have garnered sustained scientific interest for over 150 years. However, the inherently interdisciplinary nature of terpenoid research—spanning fields such as chemistry, pharmacology, and biology—poses significant challenges in integrating and communicating domain-specific knowledge across disciplines. To bridge this gap, we present TeroSeek, first by systematically extracting key scientific data and findings from terpenoid-related literature published over the past two decades to construct a curated knowledge base (KB), and then further developing an intelligent question-answering chatbot and web service powered by an AI-accelerated retrieval-augmented generation (RAG) framework. TeroSeek enables rapid access to structured, high-quality information and accurately responds to a wide range of terpenoid-related queries, demonstrating superior performance over general-purpose large language models (LLMs) in various application scenarios. Therefore, we believe that TeroSeek serves as a powerful domain-specific expert model to support the multidisciplinary terpenoid research community. The TeroSeek web service is publicly accessible at http://teroseek.qmclab.com.

**Keywords:** Terpenoid, Knowledge Base, RAG, LLM, TeroSeek

# Introduction

Terpenoids represent one of the most structurally diverse and biologically significant classes of natural products. To date, more than 150,000 distinct terpenoid molecules have been identified[1,2], supported by over 40,000 peer-reviewed publications. These compounds exhibit broad pharmacological relevance[3,4], exemplified by paclitaxel[5] (anticancer) and artemisinin[6] (antimalarial). Consequently, terpene molecules not only exhibit a wide range of biological activities but also find extensive industrial applications in fields such as food[7], daily chemicals, and flavor[8] production, the terpenoid field has maintained sustained and extensive research interest from the discovery of the first terpene molecule 150 years ago to the present day.

Terpenoid research spans multiple disciplines, including botany[9], pharmacology[3,4], analytical chemistry, and computational chemistry, among others. However, the integration of these diverse fields remains challenging, leading to significant communication barriers within the research community. For instance, botanists primarily investigate terpenoid biosynthetic pathways and novel structures, pharmacologists focus on their bioactivities, while computational chemists emphasize atom-scale reaction mechanisms. This results in fragmented terminology, methodologies, and research objectives across the field. Moreover, studying terpenoid bioactivity requires extensive biological experiments, structural elucidation relies on high-precision analytical instruments, and mechanistic insights demand computational modeling and high-throughput simulations. Each of these tasks involves steep learning curves, making interdisciplinary collaboration and research efforts rare. Therefore, consolidating terpenoid-related knowledge and breaking down disciplinary barriers would significantly accelerate progress in this field.

With the advancement of computer science, human capacity for knowledge integration has significantly improved. Large language models (LLMs)[10] have accelerated progress in numerous scientific and technological domains. However, LLMs inherently face several unresolved challenges: Knowledge Latency: LLMs are typically trained in discrete phases, meaning their knowledge remains static until the next training update, potentially leading to outdated information[11,12]. Hallucination Risks: While the probabilistic nature of LLM outputs enables extrapolation, it also introduces inaccuracies[13-15]. In critical fields, such errors could have catastrophic consequences[16]. Lack of Attribution: Even when providing correct answers, LLMs often fail to disclose knowledge sources, undermining trust in scientific applications. Implicit Knowledge Storage: All learned information is embedded implicitly within neural network parameters, necessitating increasingly larger models to encompass broader facts. This limits their ability to precisely access or manipulate knowledge at the margins[17].

Enhancing retrieval-augmented generation (RAG)[16,18-20] can effectively improve the performance of LLMs in specific tasks and has already achieved success in multiple domains[21-30]. Before answering a question, RAG retrieves reference materials, combines them with the question as part of the prompt, and then inputs this combined information to the LLM. This approach does not rely solely on the implicit knowledge encoded in the model's parameters. Instead, it significantly expands the LLM's knowledge base, enabling access to traceable and verifiable text sources without repeated training or fine-tuning. Additionally, it may effectively reduce hallucinations[31,32]. However, the improvement in LLM output brought by RAG heavily depends on the accuracy and relevance of the reference materials. If the retrieved references are inaccurate or poorly matched, RAG may not only fail to enhance the LLM's performance but could also exacerbate hallucinations. Therefore, constructing a correct and easily retrievable knowledge base is a crucial component of RAG technology.

News articles, research papers, books, reports, emails, social media posts, and other materials all contribute to building a knowledge base. However, the quality of information on the internet varies widely, while internal team information often has limited coverage. Processing and verifying information from diverse sources undoubtedly increase the difficulty in the early stages of constructing a knowledge base. In knowledge-intensive specialized fields, the authority and structure of knowledge should be prioritized. Among the many sources, academic papers undergo rigorous peer review and typically follow a fixed structure, making them highly suitable for building a knowledge base in scholarly domains. Once the initial knowledge base is established with high accuracy, subsequent information can be validated against the existing knowledge, enabling updates and enhancements to the knowledge base.

To further accelerate knowledge integration and interdisciplinary collaboration in the field of terpenoids, we applied large language models (LLMs) to dialogue tasks in this domain. We collected research literature related to terpenoids over the past 20 years and constructed a knowledge base, TeroSeek-KB, based on the abstracts, objective data, and research conclusions from these publications. Leveraging TeroSeek-KB and the open-source LLM DeepSeek-R1[33], we developed intelligent dialogue applications named TeroSeek-normal and TeroSeek-review, significantly enhancing the performance of LLMs in answering terpenoid-related questions. We have made all dialogue applications publicly accessible as web services at http://teroseek.qmclab.com.

## Materials and Methods

We conducted a literature search across three major databases - Web of Science (https://www.webofscience.com), Scopus (https://www.scopus.com), and PubMed(https://pubmed.ncbi.nlm.nih.gov) using "terpene" as the primary search term across all article types. Subsequently, LLMs were employed to read the abstracts of candidate papers to determine whether the article topics were related to terpenoids (Figure 1a). Subsequently, we collected 47,731 academic documents, which included 44,701 research papers and 3,030 literature reviews. All metadata such as journal names, authors, DOIs, and publication years were retained to generate standardized citations in responses.

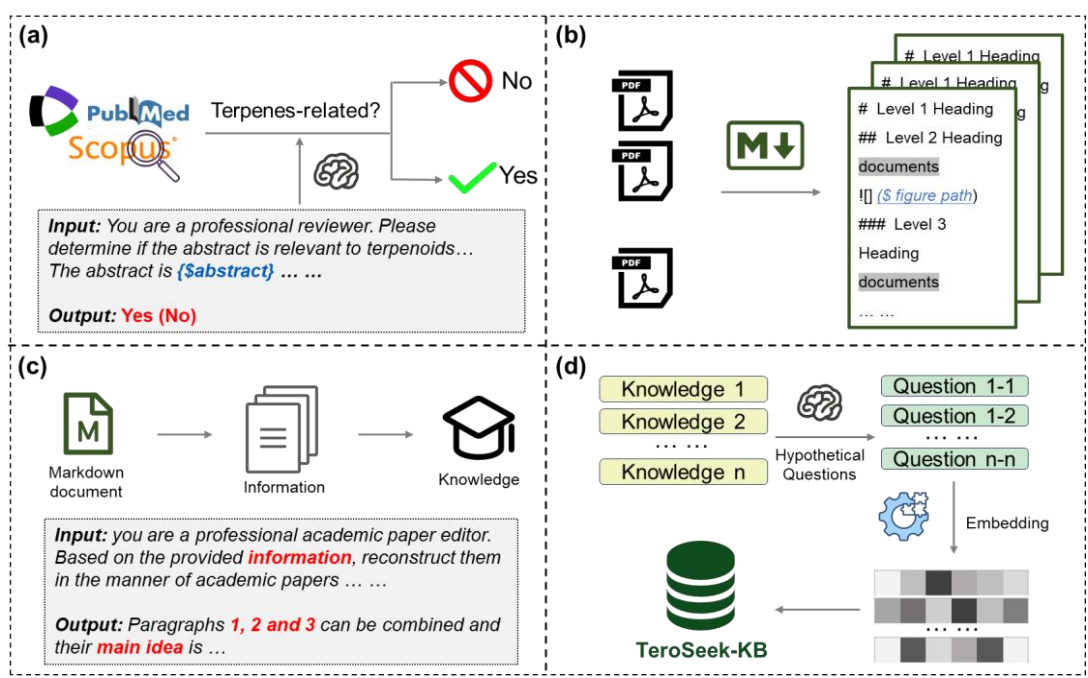

Figure 1: Data parsing and knowledge base construction process. (a) Preprocessing approach for journal articles, utilizing an LLM to screen candidate literature. (b) Conversion of PDFs to Markdown format for further parsing. (c) Employing an LLM for information cleansing and consolidation. (d) Using an LLM to generate hypothetical questions from knowledge, followed by vectorization and storage in a vector database.

The documents in pdf format were converted to Markdown format for clearer document structure. Preliminary segmentation was achieved by matching level-1 headings (prefixed with one #), level-2 headings (##), and level-3 headings (###) followed by their corresponding text (Figure 1b). After that, an LLM was used to clean and merge candidate texts, removing non-content elements (e.g., formatting symbols) and irrelevant information. For shorter or semantically similar texts, the LLM proposed merging strategies, which were implemented in subsequent processing (Figure 1c). All the Chunks will generate up to 4 hypothetical questions through LLMs, and then these questions will be converted into 2048-dimensional vectors by the embedding model and stored in the vector database (Figure 1d). Finally, 228,400 text segments with 932,577 hypothetical questions were generated to construct a knowledge base TeroSeek-KB.

The knowledge base employs a hierarchical retrieval strategy to improve both the accuracy and efficiency of searches. For each query vector, the system first searches the summary index layer—constructed from the embeddings of literature abstracts—to quickly identify key documents. This first-layer index returns up to 400 results to minimize the risk of missing relevant information. Next, candidate knowledge is further matched at the sub-chunk level. In this second layer, the system narrows down the results to a maximum of 20 entries, selecting those with a cosine similarity above 0.7 to the query vector, thus ensuring the final outputs are both concise and precise (see Figure 2a).

The backend deploys two services: an expert Q&A model and a professional research model. In the expert Q&A model, each user query is first assessed for relevance to terpenoid molecules. If relevant, the system searches the TeroKit database and returns pertinent molecular data. The user's question is then converted into a vector representation to query the article knowledge base. Initially, only the metadata of the retrieved results such as paper titles, authors, and journals is displayed in the chat interface. Next, the detailed content from these results is integrated with the user's original query to construct a prompt, which is then fed to the LLM to generate a comprehensive final answer (see Figure 2b). Due to content provider restrictions, raw article content is not included in responses, though researchers with appropriate permissions may access original URLs.

To further enhance the retrieval capabilities of the knowledge base, the research module implements a more advanced workflow. For user-specified topics, the module adopts a two-stage retrieval strategy. Initially, retrieval is conducted solely within the review-specific knowledge base to establish a comprehensive understanding of the queried topic. Based on the materials obtained in this first stage, the system generates relevant sub-questions, each of which is subsequently addressed by invoking the expert Q&A module to provide in-depth responses. Throughout the entire process, all referenced sources are systematically documented (see Figure 2c).

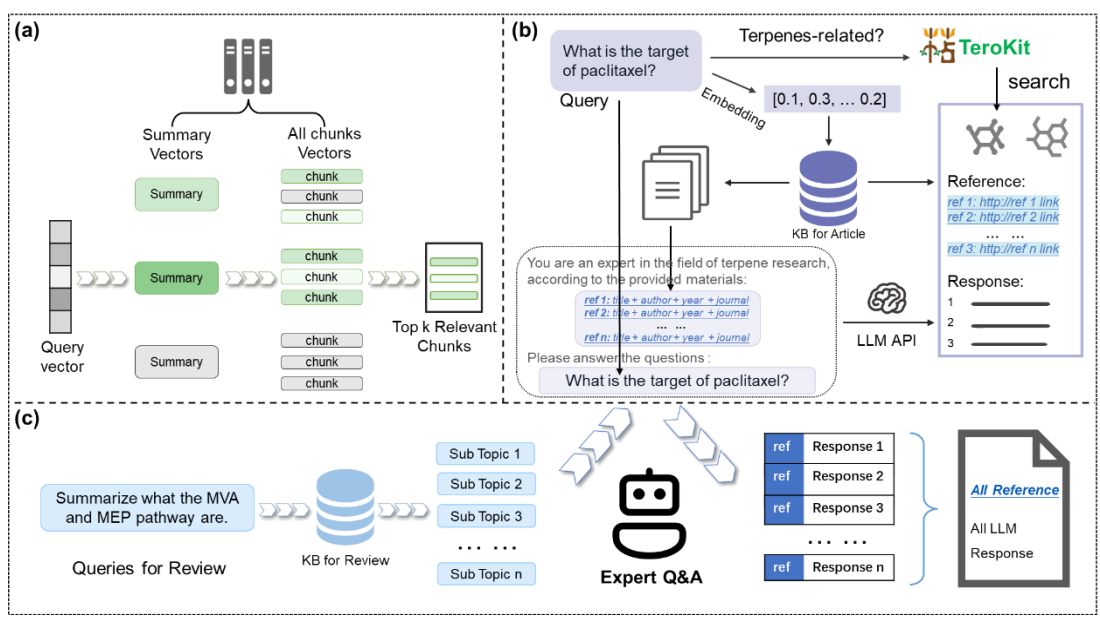

Figure 2: The model architecture of TeroSeek, (a) the vector retrieval process, (b) the response process of the Q&A model, (c) the response process of the research model.

Three top-performing open-source models as of May 2025 were utilized, including Qwen-3-235B-A22B (https://qwenlm.github.io/blog/qwen3), DeepSeek-V3[34], and DeepSeek-R1[33]. In addition, four commercial models were employed: Gemini 2.5 Pro Preview, Claude 3 Sonnet, OpenAI o3, and Grok-3 Beta. PDF parsing was conducted using Marker (https://github.com/VikParuchuri/marker). The retrieval-augmented generation (RAG) pipeline was implemented with LlamaIndex (https://www.llamaindex.ai) and Dify (https://dify.ai). Terpenoid-specific data were obtained from TeroKit (http://terokit.qmclab.com).

# Result

Based on Q&A and research models, we've developed two conversational applications: TeroSeek-normal: Enhances LLM performance for professional queries by retrieving relevant materials from the knowledge base. It academically cites all sources with clickable links to original articles, improving response interactivity. TeroSeek-review: An advanced variant that extends knowledge integration capabilities. For questions requiring more detailed answers, it first queries a review-specific knowledge base for broader context, then generates sub-questions processed iteratively by TeroSeek-normal, and finally synthesizes these responses into a comprehensive answer.

Figure 3 showcases the response of TeroSeek-normal to "What is the target of paclitaxel?" compared with its backbone LLM, DeepSeek-R1. TeroSeek-normal first retrieved relevant terpenoid molecules from TeroKit, displaying their SMILES and images with clickable names that link to the TeroKit interface for detailed information. After searching the knowledge base and identifying eight reference papers, it primarily used Ref3—a review article on paclitaxel—to formulate its comprehensive answer.

The paper explicitly states that paclitaxel shows remarkable efficacy against metastatic ovarian and breast cancers, while also demonstrating therapeutic potential for esophageal, lung, colorectal, and brain cancers, as well as lymphoma. The review explains that paclitaxel specifically targets the β-subunit of tubulin, promoting polymerization and assembly. This process depletes intracellular tubulin, disrupts cellular functions, prevents spindle formation, causes mitotic arrest at the G2/M phase, and ultimately triggers cancer cell death. Notably, unlike other antimitotic agents, paclitaxel selectively inhibits abnormal mitosis in tumor cells while sparing normal cell division, resulting in a more favorable side effect profile. This evidence directly supports the answer of TeroSeek-normal. Furthermore, Ref1, Ref2, Ref4, Ref5, Ref7, and Ref8 primarily focus on the anticancer properties of terpenoids, including paclitaxel, without detailing its mechanism of action, thus they were not utilized as direct references. Ref6, on the other hand, contains no mention of paclitaxel whatsoever—it solely addresses terpenoids in the context of breast cancer, likely resulting from a retrieval error.

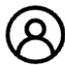

Figure 3: Response Demonstration Case - Knowledge Q&A For the question: "What is the target of paclitaxel?", the upper section displays the output from TeroSeek-normal, while the lower section presents the output from DeepSeek-R1. The illustrated outputs have been reformatted and appropriately simplified for clarity.

The TeroSeek-normal demonstrates superior capability in extracting information from tables and figures within references. For instance, it precisely cites Table 5 from Ref1, which catalogs bioactive diterpenes and their applications, specifically highlighting paclitaxel's anticancer properties. In comparison, the backbone model DeepSeek-R1 provides only broad mentions of key elements such as tubulin and mitotic inhibition without traceability to specific sources, resulting in less interpretable output. This example clearly illustrates how a knowledge base substantially enhances the precision and reliability of LLM responses.

The RAG strategy significantly improves LLMs' comprehension of scientific inquiries. When addressing real-world scientific research challenges, RAG notably elevates response quality. When users present specific operational research issues, TeroSeek-normal retrieves relevant information from multiple reference articles and delivers detailed solutions. More importantly, TeroSeek presents procedural examples from the referenced articles alongside its solutions, providing practical context.

To further leverage the literature retrieval capabilities of the TeroSeek-KB, we have developed the TeroSeek-review feature. This enhancement generates targeted sub-questions related to the user's query by referencing the review knowledge base, then iteratively employs TeroSeek-normal to produce intermediate results before synthesizing all outputs into comprehensive conclusions. This methodical approach is particularly effective for analyzing complex scientific problems.

To objectively evaluate LLM performance improvements through enhanced RAG strategies in the terpenoid domain, we developed a specialized test set comprising both AI-generated synthetic questions and manually curated expert-level questions. As illustrated in Figure 4a, the test set consists of two components: (1) AI-generated questions, where an LLM created numerous potential questions based on article passages, which were then manually screened and verified by human experts; and (2) Real-world questions collected from practical scenarios and answered by domain experts. All answers include explanations and references to knowledge sources. The final test set contains 126 multiple-choice questions, each with four options and a single correct answer. Figure 4b illustrates the disciplinary distribution of these questions.

We evaluated the three top-performing open-source models available as of May 2025: Qwen3-235B-A22B, DeepSeek-V3, and DeepSeek-R1. We assessed both their baseline performance and their enhanced capabilities when augmented with TeroSeek-KB. To ensure response consistency and reliability, each question was answered five times by each model. After excluding cases where models produced identical responses for simpler questions, we refined our test set to 41 entries, with results presented in Figure 4c and Figure 4d.

The implementation of the RAG strategy yielded substantial improvements in response accuracy. When integrated with TeroSeek-KB, the models demonstrated notable performance gains: DeepSeek-V3 improved from 0.37 to 0.66 accuracy, DeepSeek-R1 increased from 0.50 to 0.78, and Qwen2-235B-A22B rose from 0.46 to 0.74, as illustrated in Figure 4c. Based on these results, we selected DeepSeek-R1 as the backbone model for our web service, TeroSeek-Normal. We subsequently benchmarked TeroSeek-normal against other leading language models using the same test set. The comparative results showed: Gemini 2.5 Pro achieved 0.61 accuracy, Claude 3.7 Sonnet scored 0.51, o3 attained 0.46, and Grok 3 Beta reached 0.44. TeroSeek-Normal demonstrated superior performance on terpenoid-related questions, as shown in Figure 4d.

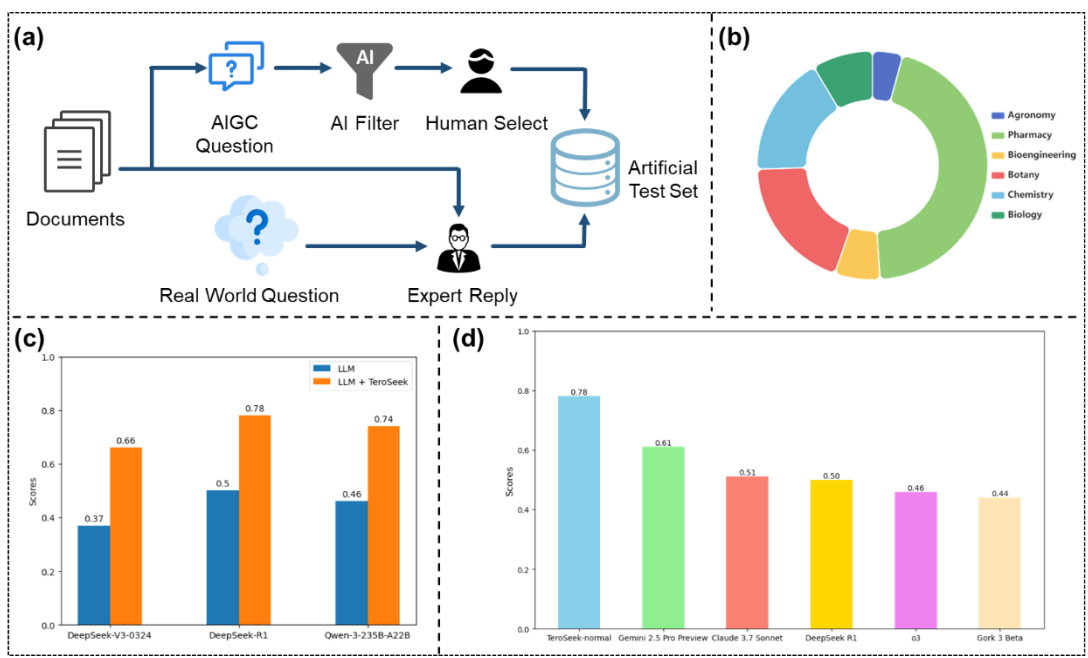

Figure 4: Benchmark of TeroSeek-normal with other leading LLMs. (a) Performance of backbone models DeepSeek-V3, DeepSeek-R1 and Qwen-3-235B-A22B on the test set and their enhanced performance with TeroSeek-KB. (b) Comparison between TeroSeek-normal and other LLMs in terms of scores. (c) Test set construction pipeline. (d) Distribution of test set questions across different disciplines.

# Discussion

In this study, the expert dialogue module generates responses by leveraging knowledge base retrieval. While this approach requires more computational resources than fine-tuned domain-specific models due to the additional input tokens needed to process retrieved knowledge, it offers significant advantages over fine-tuning methods. Fine-tuning strategies are tightly coupled with the underlying backbone model, limiting their adaptability to rapidly evolving LLM technology. In contrast, the RAG strategy provides greater flexibility and can more readily incorporate advances in foundation models.

Model Architecture: The key advantage of this approach is the separation of the knowledge base, question-answering pipeline, and backbone LLM. Unlike fine-tuning methods that are wholly dependent on the backbone LLM's performance, our architecture offers greater flexibility. When a backbone LLM is updated or a superior model emerges, fine-tuning and parameter optimization typically require starting from scratch. Our decoupled design eliminates this limitation. Furthermore, fine-tuning existing LLMs with restricted training datasets heightens the risk of overfitting, compromising model generalizability. Our strategy externally stores all knowledge in a dedicated repository, effectively expanding the LLM's knowledge boundaries while preserving its core capabilities.

Data Updates: Crucially, domain knowledge undergoes constant evolution, with newer information typically being more timely and relevant. Updating knowledge in fine-tuned models presents significant challenges because such models encode knowledge within their parameters. Any parameter adjustment necessitates comprehensive model re-evaluation to prevent overfitting. In contrast, our RAG strategy based on an external knowledge base simplifies updates—new knowledge merely requires integration into the repository. Model iteration becomes more streamlined as well, involving only the replacement of the LLM provider's API in the pipeline or local deployment of an updated model.

Training Data Construction: Supervised fine-tuning approaches predominantly rely on question-answer paired dialogue data. Many existing methods generate "pseudo" question-answer datasets using LLMs, which often fail to accurately represent actual user queries. Our current strategy circumvents this process entirely, requiring only the annotation of new knowledge to ensure its retrievability by the LLM.

Our workflow began by compiling essential terpenoid literature into a comprehensive knowledge base, complemented by efficient vector indexes for swift and accurate retrieval. We've incorporated data from over 40,000 literature abstracts and conducted thorough full-text analysis of research papers, ensuring no detail was overlooked. The system surpasses existing LLMs in answering terpenoid-specific queries while maintaining academic integrity through citation-attached responses that ensure full traceability. Users benefit from an intuitive interface featuring clickable references and interactive functionality.

TeroSeek was deliberately designed for adaptability, accommodating the rapid evolution of LLM technology and AI advancements. Rather than embedding knowledge within model parameters, we implemented the more flexible and scalable RAG framework. Our architecture separates each component—knowledge updates, knowledge base, vector operations, Q&A agent, and web services—into independent modules that communicate via network protocols, ensuring minimal coupling. This modular approach allows for targeted optimization of any subsystem, yielding immediate performance improvements. The decoupled knowledge base and update pipeline enable real-time data integration. The standalone web service and Q&A agent further support service expansion without compromising core functionality.

In conclusion, this study established TeroSeek-KB, a specialized terpenoid domain knowledge base built from comprehensive literature data. By integrating this knowledge base with LLMs through a RAG pipeline, we significantly enhanced knowledge coverage and response quality in the terpenoid domain. As LLMs continue to evolve and our knowledge base expands, the accuracy of terpene-related responses will steadily improve. The TeroSeek web service serves as both a unified resource hub for terpenoid research and a communication bridge for researchers from diverse backgrounds, ultimately facilitating deeper exploration of specialized questions in the field.

# Conclusion

This study pioneers the application of RAG technology in the terpenoid field, creating a comprehensive knowledge base from scholarly literature and launching an open-access web platform. The TeroSeek system significantly enhances LLM output interpretability and demonstrates superior performance in addressing terpenoid-specific queries. This innovation promises to catalyze interdisciplinary collaboration in terpenoid research. In the future, TeroSeek plans to expand its knowledge base to incorporate books, online forums, encyclopedic content, and diverse reference materials. A key priority will be strengthening integration with major database platforms, enabling researchers to access terpenoid molecules, enzymes, and genetic information with unprecedented precision and efficiency.

# Reference

1   Zeng, T. *et al.* TeroKit: A Database-Driven Web Server for Terpenome Research. *Journal of Chemical Information and Modeling* **60**, 2082-2090, doi:10.1021/acs.jcim.0c00141 (2020).
2   Zeng, T., Chen, Y., Jian, Y., Zhang, F. & Wu, R. Chemotaxonomic investigation of plant terpenoids with an established database (TeroMOL). **235**, 662-673, doi:https://doi.org/10.1111/nph.18133 (2022).
3   Cox-Georgian, D., Ramadoss, N., Dona, C. & Basu, C. in *Medicinal Plants: From Farm to Pharmacy*   (eds Nirmal Joshee, Sadanand A. Dhekney, & Prahlad Parajuli)   333-359 (Springer International Publishing, 2019).
4   Jaeger, R. & Cuny, E. Terpenoids with Special Pharmacological Significance: A Review. **11**, 1934578X1601100946, doi:10.1177/1934578x1601100946 (2016).
5   Singla, A. K., Garg, A. & Aggarwal, D. J. I. j. o. p. Paclitaxel and its formulations. **235**, 179-192 (2002).
6   White, N. J. J. S. Qinghaosu (artemisinin): the price of success. **320**, 330-334 (2008).
7   José, S. C. *et al.* Plant-Derived Terpenoids: A Plethora of Bioactive Compounds with Several Health Functions and Industrial Applications—A Comprehensive Overview. *Molecules* **29**, 3861, doi:10.3390/molecules29163861 (2024).
8   Patel, T., Ishiuji, Y. & Yosipovitch, G. J. J. o. t. A. A. o. D. Menthol: a refreshing look at this ancient compound. **57**, 873-878 (2007).
9   Christopher, I. K. & Jörg, B. Plant Terpenoids. *Wiley Encyclopedia of Chemical Biology*, 1-10, doi:10.1002/9780470048672.wecb596 (2008).
10  Vaswani, A. *et al.* Attention is all you need. *Advances in neural information processing systems* **30** (2017).
11  Lewis, P. *et al.* Retrieval-augmented generation for knowledge-intensive nlp tasks. *Advances in neural information processing systems* **33**, 9459-9474 (2020).
12  He, H., Zhang, H. & Roth, D. Rethinking with retrieval: Faithful large language model inference. *arXiv preprint arXiv:2301.00303* (2022).
13  Huang, Y. *et al.* in *International Conference on Machine Learning.*   20166-20270 (PMLR).
14  Raunak, V., Menezes, A. & Junczys-Dowmunt, M. The curious case of hallucinations in neural machine translation. *arXiv preprint arXiv:2104.06683* (2021).

15 Ji, Z. *et al.* Survey of hallucination in natural language generation. *ACM computing surveys* **55**, 1-38 (2023).

16 Chen, J., Lin, H., Han, X. & Sun, L. in *Proceedings of the AAAI Conference on Artificial Intelligence.* 17754-17762.

17 Khattab, O. *et al.* Demonstrate-search-predict: Composing retrieval and language models for knowledge-intensive nlp. *arXiv preprint arXiv:2212.14024* (2022).

18 Jiang, Z. *et al.* in *Proceedings of the 2023 Conference on Empirical Methods in Natural Language Processing.* 7969-7992.

19 Guu, K., Lee, K., Tung, Z., Pasupat, P. & Chang, M. in *International conference on machine learning.* 3929-3938 (PMLR).

20 Fan, W. *et al.* in *Proceedings of the 30th ACM SIGKDD Conference on Knowledge Discovery and Data Mining.* 6491-6501.

21 Liu, S. *et al.* Multi-modal molecule structure–text model for text-based retrieval and editing. **5**, 1447-1457 (2023).

22 Xu, J., Crego, J.-M. & Senellart, J. in *Annual Meeting of the Association for Computational Linguistics.* 1570-1579 (Association for Computational Linguistics).

23 Long, C., Liu, Y., Ouyang, C. & Yu, Y. J. a. p. a. Bailicai: A Domain-Optimized Retrieval-Augmented Generation Framework for Medical Applications. (2024).

24 Yang, Z. *et al.* ShennongAlpha: an AI-driven sharing and collaboration platform for intelligent curation, acquisition, and translation of natural medicinal material knowledge. **11**, 32 (2025).

25 Lozano, A., Fleming, S. L., Chiang, C.-C. & Shah, N. in *PACIFIC SYMPOSIUM ON BIOCOMPUTING 2024.* 8-23 (World Scientific).

26 Sun, B. *et al.* PanKB: An interactive microbial pangenome knowledgebase for research, biotechnological innovation, and knowledge mining. *Nucleic Acids Research* **53**, D806-D818, doi:10.1093/nar/gkae1042 %J Nucleic Acids Research (2024).

27 Yeganova, L. *et al.* LitSense 2.0: AI-powered biomedical information retrieval with sentence and passage level knowledge discovery. *Nucleic Acids Research*, doi:10.1093/nar/gkaf417 (2025).

28 Wang, J. *et al.* CAUSALdb2: an updated database for causal variants of complex traits. *Nucleic Acids Research* **53**, D1295-D1301, doi:10.1093/nar/gkae1096 %J Nucleic Acids Research (2024).

29 Members, C.-N. & Partners. Database Resources of the National Genomics Data Center, China National Center for Bioinformation in 2025. *Nucleic Acids Research* **53**, D30-D44, doi:10.1093/nar/gkae978 %J Nucleic Acids Research (2024).

30	Zhang, R. *et al.* PlantGPT: An Arabidopsis-Based Intelligent Agent that Answers Questions about Plant Functional Genomics.  **n/a**, e03926, doi:https://doi.org/10.1002/advs.202503926.
31	Shuster, K., Poff, S., Chen, M., Kiela, D. & Weston, J. J. a. p. a. Retrieval augmentation reduces hallucination in conversation.   (2021).
32	Xu, J., Szlam, A. & Weston, J. J. a. p. a. Beyond goldfish memory: Long-term open-domain conversation.   (2021).
33	Guo, D. *et al.* Deepseek-r1: Incentivizing reasoning capability in llms via reinforcement learning.   (2025).
34	Liu, A. *et al.* Deepseek-v3 technical report.   (2024).